\documentclass[reprint,superscriptaddress,aip,apl]{revtex4}
\pdfoutput=1
\usepackage[utf8]{inputenc}
\usepackage{natbib}
\usepackage{amsmath}
\usepackage{amsfonts}
\usepackage{amssymb}
\usepackage{lmodern}
\usepackage{graphicx}
\usepackage[export]{adjustbox}  
\usepackage{tabularx}
\usepackage{hhline}
\usepackage{multirow}
\usepackage{dcolumn}
\usepackage{longtable} 
\usepackage{placeins} 
\usepackage[color=orange!20]{todonotes} 
\usepackage{ulem} 
\usepackage{booktabs}
\usepackage[separate-uncertainty = true,multi-part-units=single]{siunitx}
\usepackage{color}
\usepackage{nicefrac}
\usepackage{glossaries}
\usepackage{layouts} 

\newcommand{\bfs}{$\rm BaFe_2Se_3$}

\definecolor{gray}{gray}{0.67}
\definecolor{Gray}{gray}{0.45}

\newcommand{\vio}[1]{\textcolor{violet!80}{#1}}

\newcommand{\ill}{Institut Laue-Langevin, 71 av. des Martyrs, 38000 Grenoble, France}
\newcommand{\neel}{Institut N\'{e}el, CNRS, 25 av. des Martyrs, 38042 Grenoble, France}
\newcommand{\UGA}{Universit\'{e} Grenoble Alpes}
\newcommand{\LPS}{Université Paris-Saclay, CNRS, Laboratoire de Physique des Solides, 91405, Orsay, France}
\newcommand{\SOLEIL}{Synchrotron SOLEIL, L\'\ Orme des Merisiers, Saint Aubin BP 48, 91192, Gif-sur-Yvette, France}
\newcommand{\SPEC}{SPEC, CEA, CNRS-UMR3680, Université Paris-Saclay, Gif-sur-Yvette Cedex 91191, France}

\newcolumntype{d}[1]{ D{.}{.}{#1} }

\begin{document}
\newcolumntype{d}[1]{ D{.}{.}{#1} }
\title{Space Group Symmetry of BaFe$_2$Se$_3$:\\ \textit{ab initio}-Experiment Phonon Study}

\author{M.~J. Weseloh}
\email{Maria.Weseloh@neel.cnrs.fr}
\affiliation{\neel,~\UGA}
\affiliation{\ill}

\author{V.~Bal\'{e}dent}
\affiliation{\LPS}

\author{W.~Zheng}
\affiliation{\LPS}

\author {M.~Verseils}
\affiliation{\SOLEIL}

\author {P.~Roy}
\affiliation{\SOLEIL}

\author {J.~B.~Brubach}
\affiliation{\SOLEIL}

\author{D.~Colson}
\affiliation{\SPEC}

\author{A.~Forget}
\affiliation{\SPEC}

\author{P.~Foury-Leylekian}
\affiliation{\LPS}

\author{M.-B.~Lepetit}
\email{Marie-Bernadette.Lepetit@neel.cnrs.fr}
\affiliation{\neel,~\UGA}
\affiliation{\ill}

\begin{abstract}
  This paper presents a study of the structure dynamics in \bfs. We combined
   first-principle calculations, infrared measurements and a
  thorough symmetry analysis.  Our study confirms that $Pnma$ cannot be the space group of
  {\bfs}, even at room temperature. The phonons assignment requires
  $Pm$ to be the \bfs\ space group, not only in the magnetic phase, but also in
  the paramagnetic phase at room temperature. This is due to a strong coupling
  between a short range spin-order along the ladders, and the lattice degrees of freedom
  associated with the Fe-Fe bond length. This coupling induces a change in the bond-length
  pattern from an alternated trapezoidal one (as in $Pnma$) to an alternated
  small/large rectangular one. Out of the two patterns, only the latter is fully compatible with the observed block-type magnetic
  structure.  Finally\vio{,} we propose a complete symmetry analysis of the \bfs\
  phase diagram in the 0-600\,K range.
\end{abstract}

\maketitle

\section{Introduction}
\label{Intro}

Strong quantum entanglement of electron wave-functions, also called strong
electronic correlation, is at the origin of many remarkable properties. This
is the case, for instance, with superconductivity, colossal
magneto-resistance, or magneto-electric coupling. Since electronic
correlations are intrinsic quantum mechanical effects, the related properties
are primarily expected at low temperatures. However, a few families of
compounds exhibit exceptionally high working temperatures. An example is given
by the cuprates family~\cite{Supra_Muller_86} with its record superconducting
transition temperature of 164~K under 45~GPa
pressure~\cite{Cu-164K-pression-Gao1994}, and 133~K at ambient
pressure~\cite{Cu-133K-Hg-Schilling1993}.  Another family, in which
high-temperature superconductivity was
found~\cite{Pnicture-decouverte-Takahashi2008}, is the iron-based pnictides
family. In this family the inherent multi-orbital character adds a lot of
complexity, therefore, pnictides have been the subject of numerous studies
since their discovery in 2003~\cite{Pnicture-decouverte-2006}. In addition to
their superconducting properties, some of the pnictides exhibit multiferroic
properties~\cite{FE-Zhang2016} (a state with at least two coexisting/coupled
ferroic/antiferroic orders), thus increasing their interest for the community.

In the last decade, a lot of research effort has been devoted to
multiferroicity, and more specifically to magneto-electric multiferroics. Due
to the coupled nature of their electrical and magnetic orders,
magneto-electric (ME) compounds offer the possibility to control their
magnetic properties by applying a simple voltage, or control their
polarization or dielectric constant by applying a magnetic field.  Hence, they
are highly promising for new kinds of electronic devices.

Among the ME compounds, \bfs\ is one of the few ME materials exhibiting both a
magnetic order and a polarization at high temperatures.  Indeed, below the
N\'{e}el temperature ($T_N$) a long-range block-type antiferromagnetic
ordering sets in~\cite{Liu_2019,Krzton_Maziopa,Popovic,Caron2011,Gao}.  $T_N$
is reported in a quite large temperature range, from $T_N\simeq 230\rm~K$ to
$256\rm~K$, according to the different authors. Gao {\em et al.} have
attributed the discrepancies to different stoichiometric ratios in grown
crystals~\cite{Gao}.  Additionally to the large $T_N$, \bfs\ shows
quasi-one-dimensional superconductivity under
pressure~\cite{takahashi_pressure_2015}.

X-ray experiments first proposed the non-polar space group
$Pnma$~\cite{Krzton_Maziopa} for \bfs. A few years later, Dong {\em et
  al.}~\cite{Dagotto2014} theoretically predicted that the observed block-type
magnetic order should induce a polar symmetry lowering, due to
exchange-striction effects.  Weak intensity was later observed on the $hk0$,
$h=2n+1$ peaks of the X-ray diffraction pattern, that are forbidden in the
$Pnma$ group~\cite{Zheng}.  Instead of $Pmna$, the polar space group $Pmn2_1$
was proposed for \bfs's room temperature structure~\cite{Zheng}. Below
$T_N$, a further symmetry lowering was observed and assigned to a transition
from $Pmn2_1$ to $Pm$~\cite{Zheng}.

\begin{figure}[h]
a) \resizebox{5.cm}{!}{\includegraphics[valign=t]{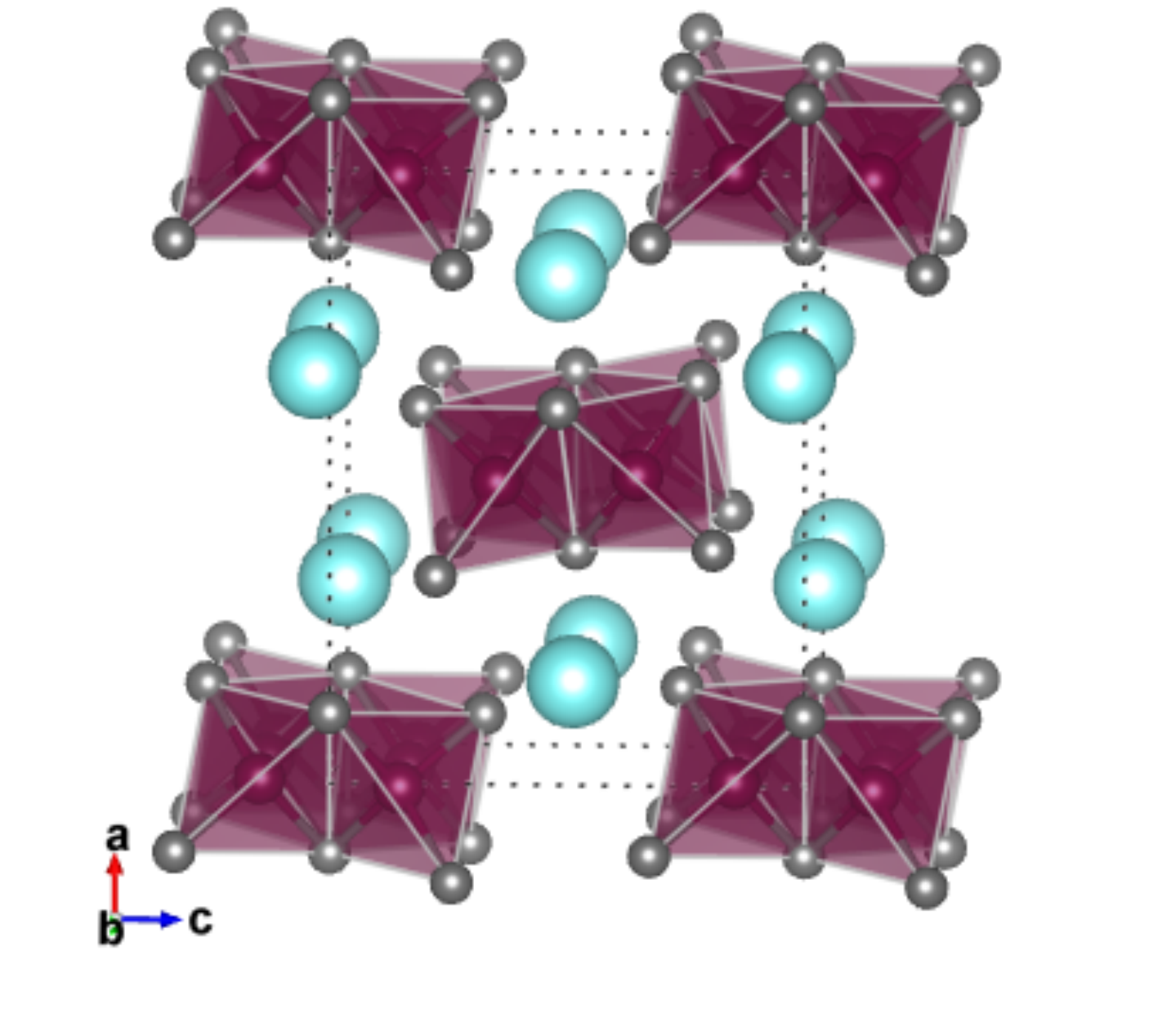}} 
b) \resizebox{2.2cm}{!}{\includegraphics[valign=t]{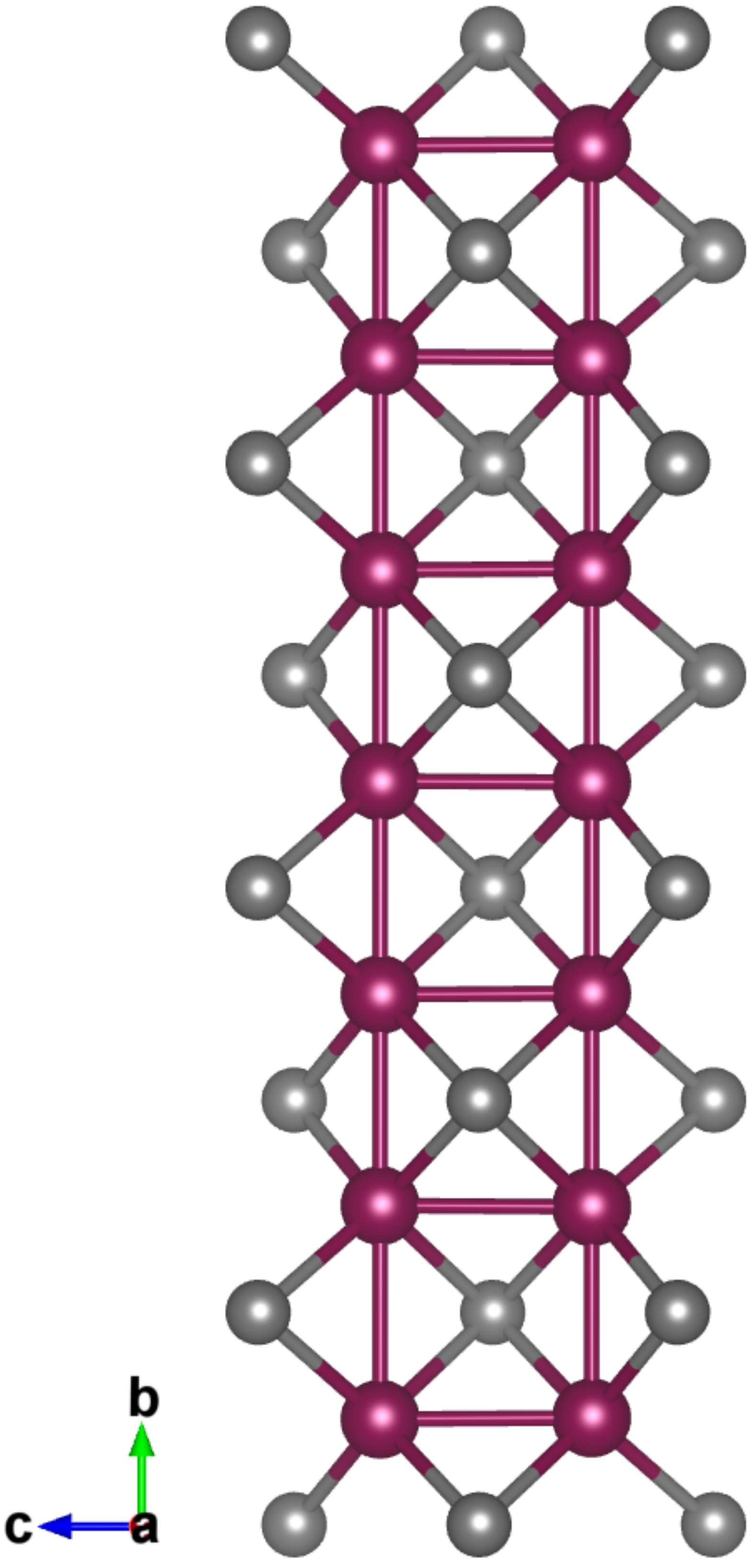}}
\caption{(a) Crystal structure  of {\bfs}. (b) Structure of the Fe-ladders.}
\label{fig:struct}
\end{figure}

The \bfs\ compound presents a quasi-one-dimensional ladder geometry coherent
with its superconducting properties (see Fig.~\ref{fig:struct}). The iron
atoms are in a $\rm FeSe_4$ tetrahedral environment. These edge-sharing
tetrahedra form two-legs ladders along the $\vec b$ direction (standard $Pnma$
setting). Each unit cell contains two ladders (one at the center and one at
the corners of the unit cell) separated by barium atoms.  In each unit cell,
there are two Fe atoms per chain along the ladder direction, and thus two
types of Fe--Fe bonds: one within the unit cell and one between cells. While
in the $Pnma$ group there is only one independent Fe site, in $Pmn2_1$, the
two chains in a ladder are associated to independent Fe sites, whilst the two
ladders in the unit cell remain symmetry related. This remaining symmetry is
however lifted at the magnetic-ordering transition with the $Pmn2_1$ to $Pm$
symmetry lowering. Another important difference between the $Pnma$, and
$Pmn2_1$ or $Pm$ structures, is the Fe--Fe bond alternation along the ladder
direction.  Whereas the $Pnma$ space group symmetry imposes that a long bond
faces a short one in the ladders, thus building alternated trapezoidal blocks
along the ladder direction, in the lower symmetry groups, this constraint is
lifted and the $Pmn2_1$/$Pm$ structures~\cite{Zheng} posses long (short) bonds
that are at the same level along the ladder direction (see
  Fig.~\ref{fig:bonds}). The latter geometry is in good agreement with the block magnetic
order seen in neutron scattering measurements~\cite{Krzton_Maziopa,Nambu2012}.
In this respect, \bfs\ is different from most multiferroics. Prominent is that
\bfs\ is rather ferrielectric than ferroelectric, with a strong polarization
in each ladder, mostly canceling out between the ladders.  Its fully
ferroelectric state has an energy predicted only slightly higher than the
ferrielectric one and holds a giant improper polarization predicted to be
$\sim 2–3\mu C/cm^2$, while the ferrielectric one was predicted to be
$\sim 0.2\mu C/cm^2$~\cite{Dagotto2014} and experimentally estimated around
$\sim 0.6\mu C/cm^2$~\cite{Du2020}.

In this paper we present a lattice dynamic study, combining experimental
infrared (IR) spectroscopy studies with first principle density-functional theory
(DFT) calculations.  It is well known that phonon spectra are of crucial
importance in multiferroic systems. In particular, they are often much more
efficient than diffraction methods to reveal weak symmetry breaking and to
distinguish between space groups.  The next section will detail both
experimental and numerical methods. Section~\ref{sec:IR} will be devoted to
the infrared measurements, and section~\ref{sec:dft} to DFT
results and discussion.

\section{Methods}
\label{Methods}

\subsection{Experimental}
\label{Methods: Exp.}
The experimental phonon modes were measured on high quality
single-crystals. Infrared spectroscopy measurements were carried out on the
AILES beamline of the SOLEIL synchrotron, with a Bruker IFS125 Michelson 
interferometer~\cite{Roy2006} equipped with a closed circle He-gas cryostat, a 4.2\,K bolometer, and a 6\,$\mu$m beam splitter for a resolution of 2\,cm$^{-1}$.
The synchrotron radiation beam was linearly polarized thanks to polyethylene polarizers.
Two pieces of the crystal used in
Ref.~\onlinecite{Zheng} were pre-aligned, in order to put the electric and
magnetic fields (labeled $(e,h)$) on the incident beam, in the
$(\vec c,\vec b)$ and $(\vec b,\vec c)$ crystallographic directions. For each
crystal, the reflectivity was recorded for several temperatures from 10 to
300~K. The absolute
reflectivity of the sample was obtained by using as reference the same gold-coated sample obtained by in situ
gold coating evaporation technique.  As a
  consequence,
each reflectivity spectra is the ratio between the reflected intensity
on the sample and the reflectivity from the gold deposited on the sample
surface.

\subsection{Theoretical}
\label{Methods: Theo.}
The phonon modes were calculated using DFT, after
a full geometry optimization within the constraints of a given space group. We
used the CRYSTAL code~\cite{crystal17-1,crystal17-2} which fully implements
the 230 space group symmetries.  Therefore, it can decipher the phonons
spectra issued from different space groups. This code also offers the
advantage to use atomic Gaussian basis sets and hence enables the use of
hybrid functionals with nearly no additional computational costs.  Since our
system is strongly correlated and presents metallic, semiconducting
and magnetically ordered phases~\cite{Liu_2019}, it is important to use a
hybrid functional in order to describe at the best the system's
electronic structure. For this purpose we use the B3LYP
functional~\cite{Becke1,Stephens}.

The atomic basis sets were chosen as all-electrons, valence 3$\zeta$+p basis
set for the Fe and Se atoms~\cite{heyd}, a relativistic core pseudo-potential
of the Stuttgart group~\cite{PseudoR13} for the Ba atom, and the associated
basis set adapted to solid-state calculations~\cite{heyd}.  Since the
unit cell does not change in the different groups, we used a
$5\times 10\times 6$~Monkhorst-Pack $\mathbf{k}$-grids ($Pnma$ axes) for all
calculations in the single unit cell, and an equivalent grid spacing for the
calculations in the double unit cell.

The phonon modes were computed at the center of the Brillouin zone using the
harmonic approximation. We performed two types of calculations: calculations
without spin polarization within a single unit cell, and calculations with
spin polarization along the chains --~$(0,\frac{1}{2},0)$ propagation vector
in the $Pnma$ axes~-- within a double unit cell. The experimental propagation
vector is
$(\frac{1}{2},\frac{1}{2},\frac{1}{2})$~\cite{Krzton_Maziopa,Nambu2012}. Hence,
to completely account for the magnetic order the calculation should be done in
a $2a \times 2b \times 2c$ supercell. Unfortunately, such a unit cell is too
large for performing geometry optimization and phonons calculations with our
current computer resources.  We thus opted for a feasible compromise that sets
properly the magnetic order within the ladders, as it is associated with the
largest magnetic integrals of the system.

\section{Infrared measurements}
\label{sec:IR}

\begin{figure}[h!]
  \centering
 a) \resizebox{7cm}{!}{\includegraphics[valign=t]{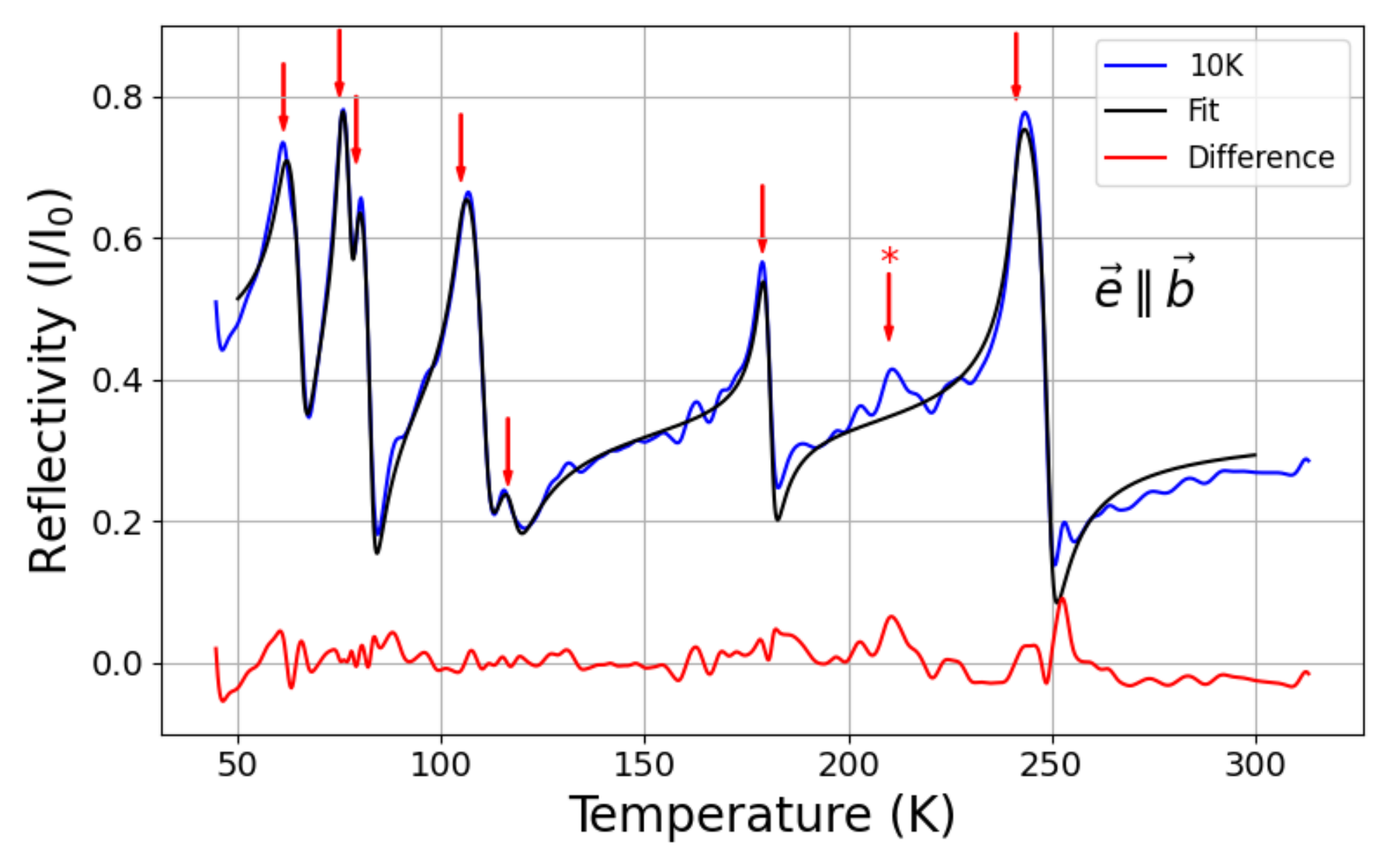}} \\
 b) \resizebox{7cm}{!}{\includegraphics[valign=t]{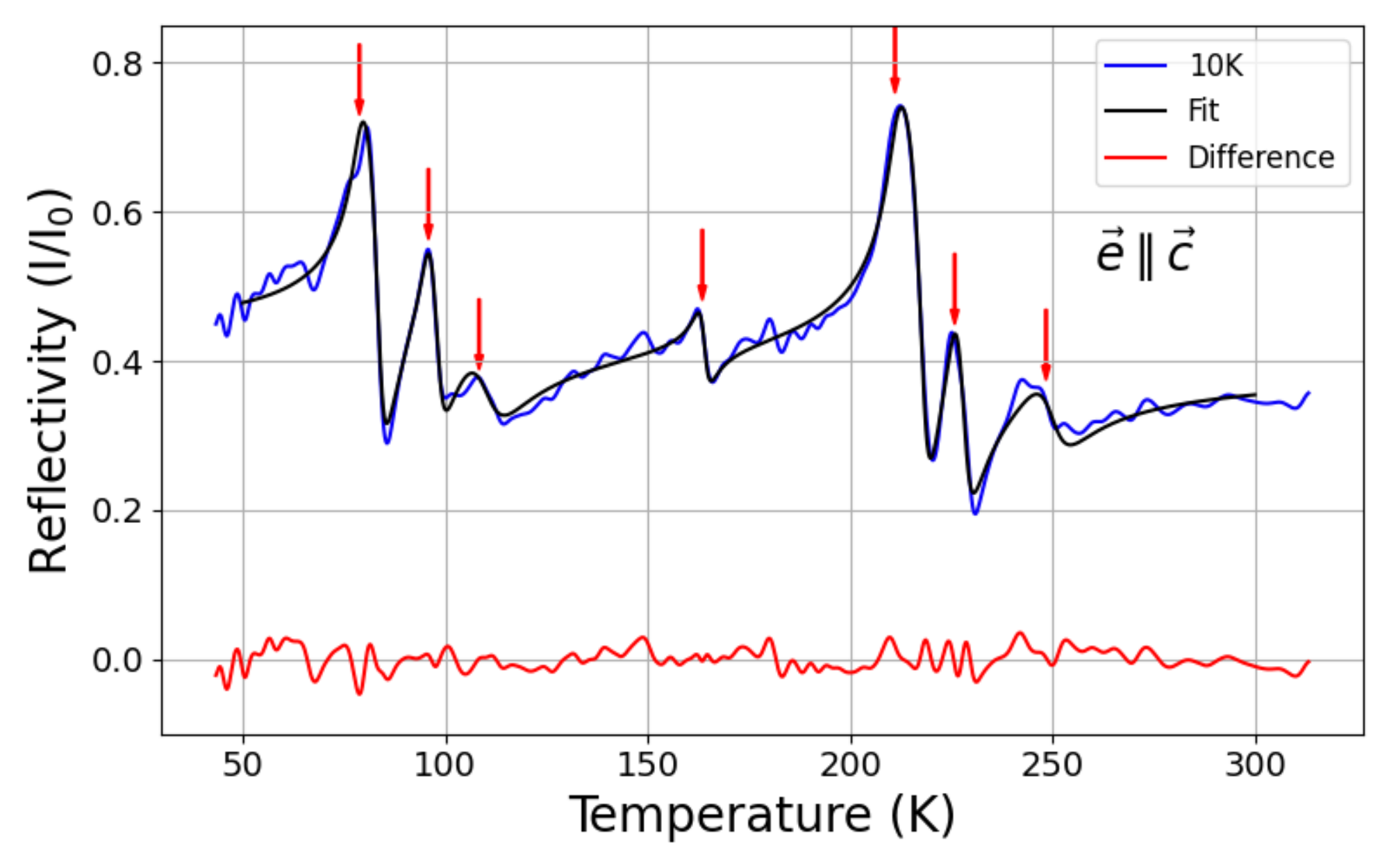}} 
\caption{Infrared measurements at 10~K.  Axes defined in the $Pnma$ standard setting, (a)
  the electric field is along the $\vec b$ direction, (b) the electric field is along the $\vec c$
  direction.  Red arrows points the phonons positions $\omega_k$, the star indicates the phonon position not successfully fitted.}
\label{fig:IR10K}
\end{figure}

\begin{figure}[h!]
  \centering
 a) \resizebox{7cm}{!}{\includegraphics[valign=t]{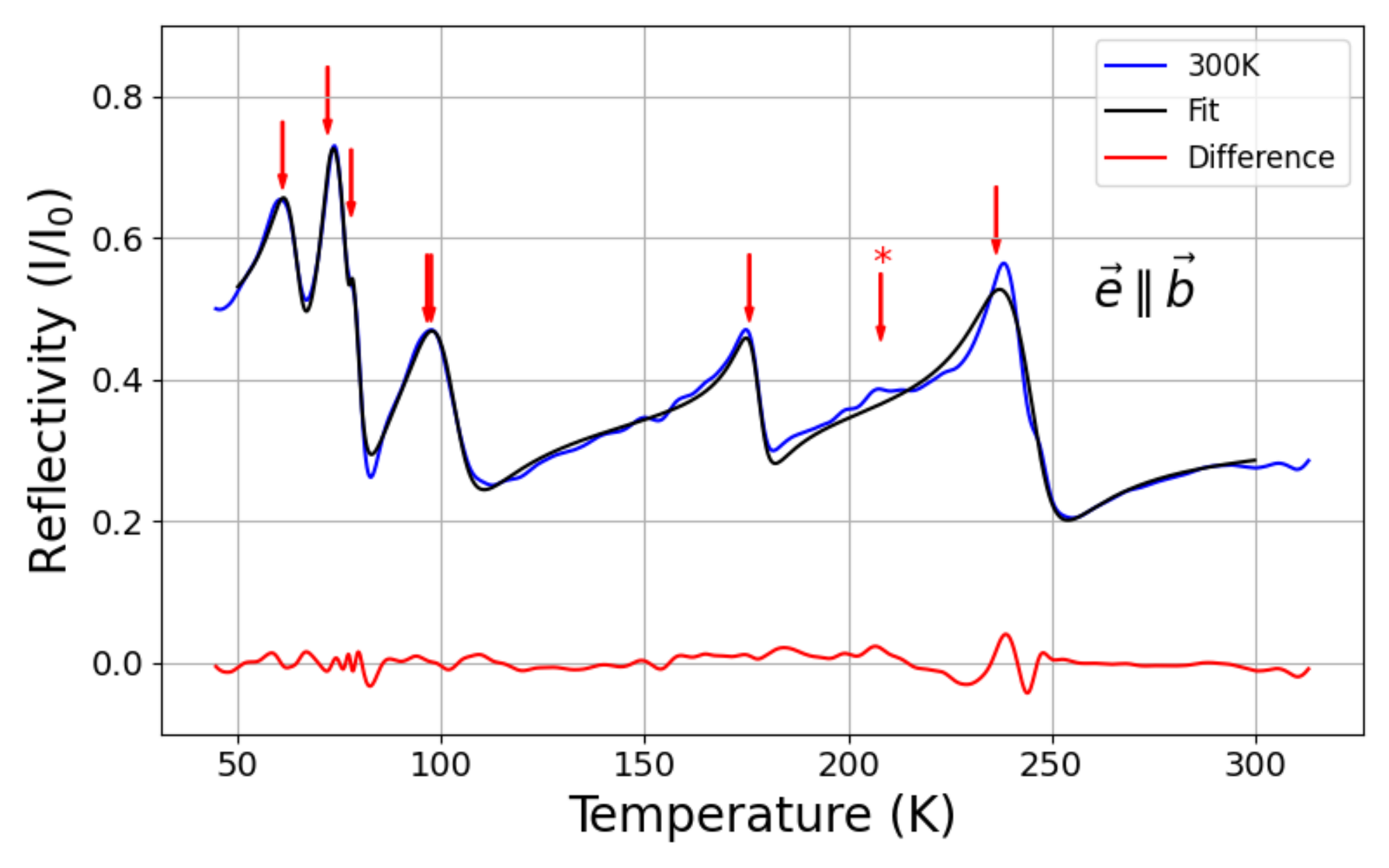}} \\
 b) \resizebox{7cm}{!}{\includegraphics[valign=t]{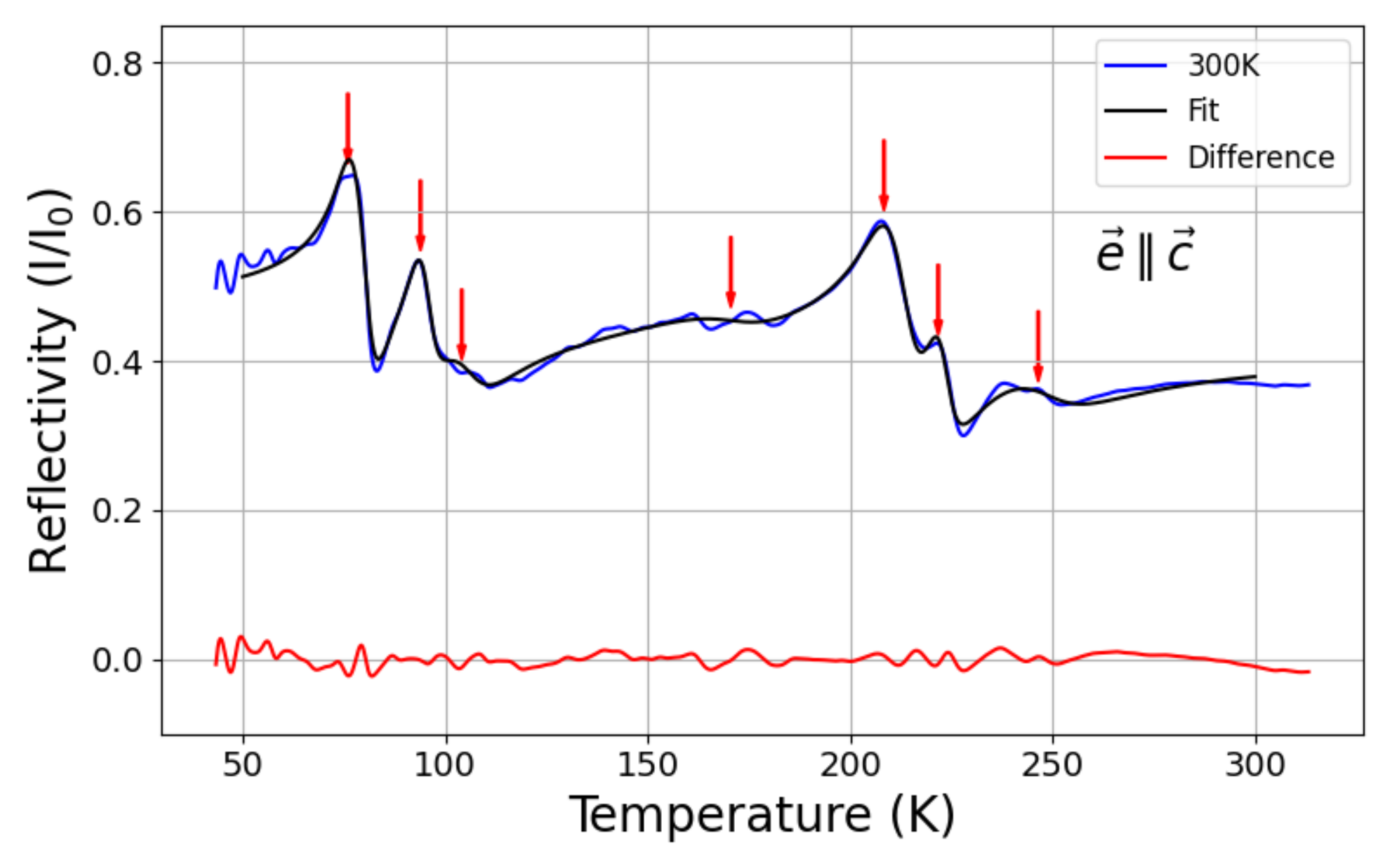}} 
\caption{Infrared measurements at 300~K.  In the $Pnma$ standard setting, (a) the electric
  field is along the $\vec b$ direction, (b) the electric field is along the $\vec c$
  direction. Red arrows points the phonons positions $\omega_k$ , the star indicates the phonon position not successfully fitted.}
\label{fig:IR300K}
\end{figure}

\begin{figure*}
  \centering
  \resizebox{0.32\linewidth}{!}{\includegraphics{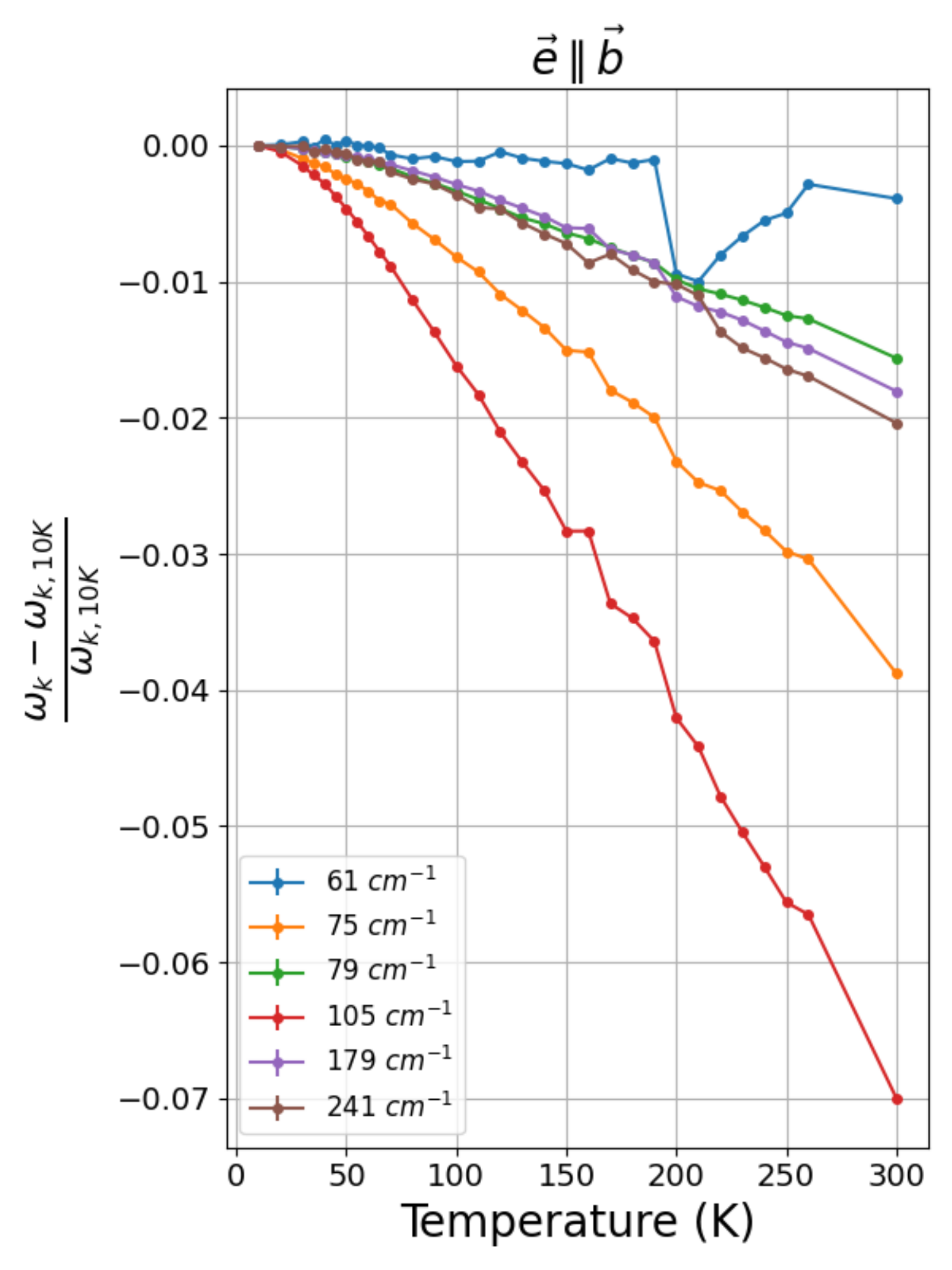}}
  \resizebox{0.32\linewidth}{!}{\includegraphics{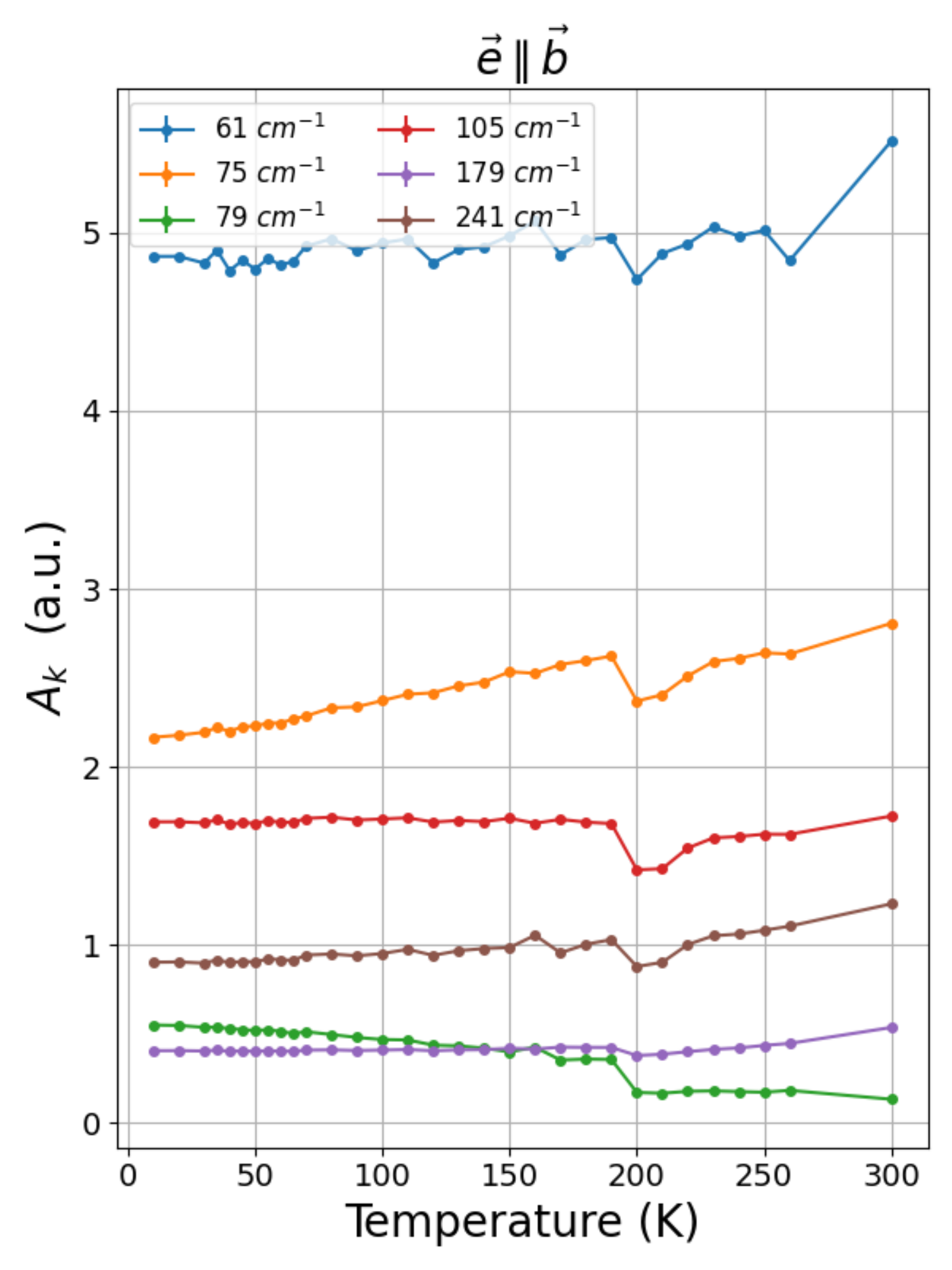}}
  \resizebox{0.32\linewidth}{!}{\includegraphics{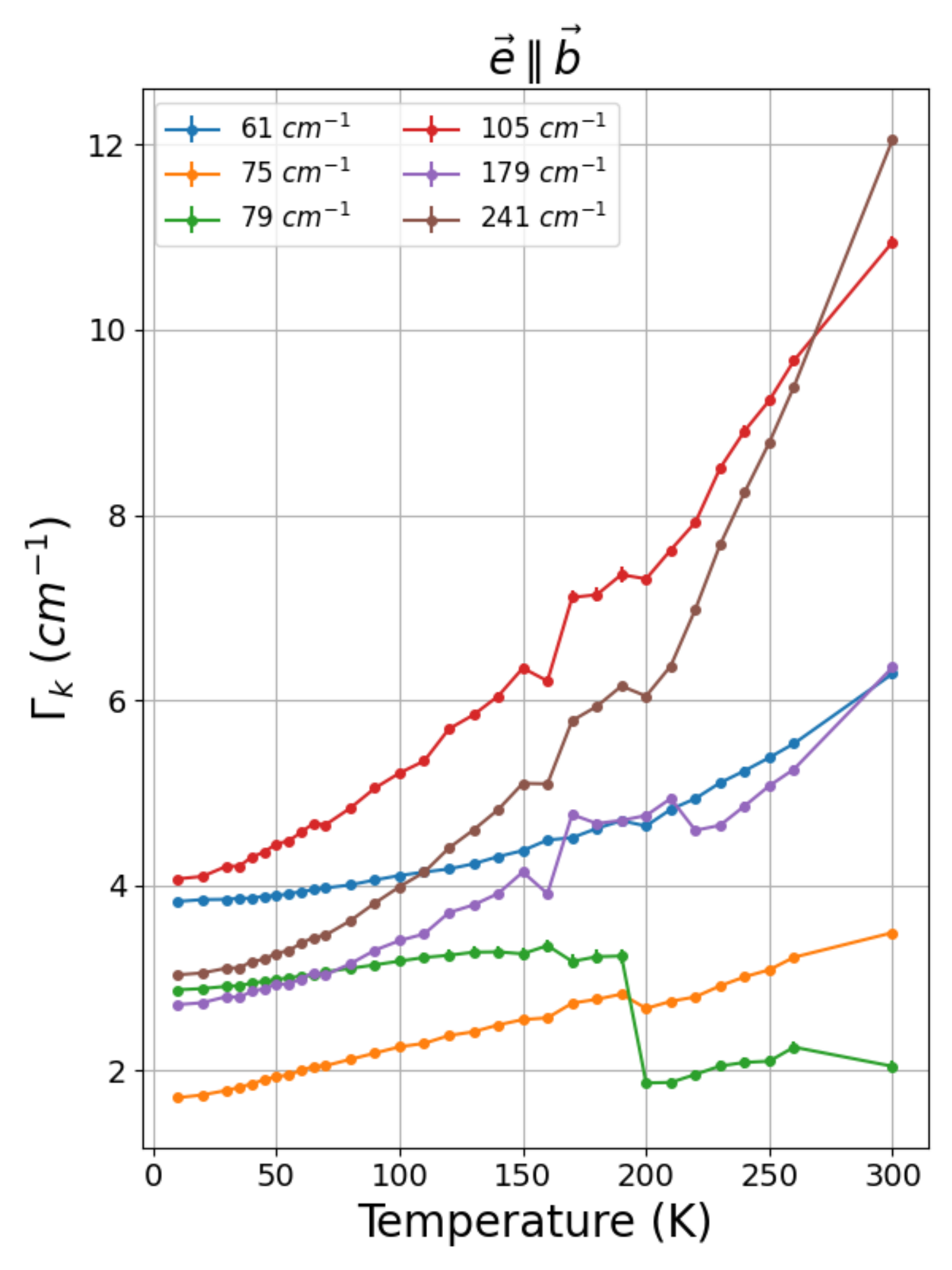}} \\
  \resizebox{0.32\linewidth}{!}{\includegraphics{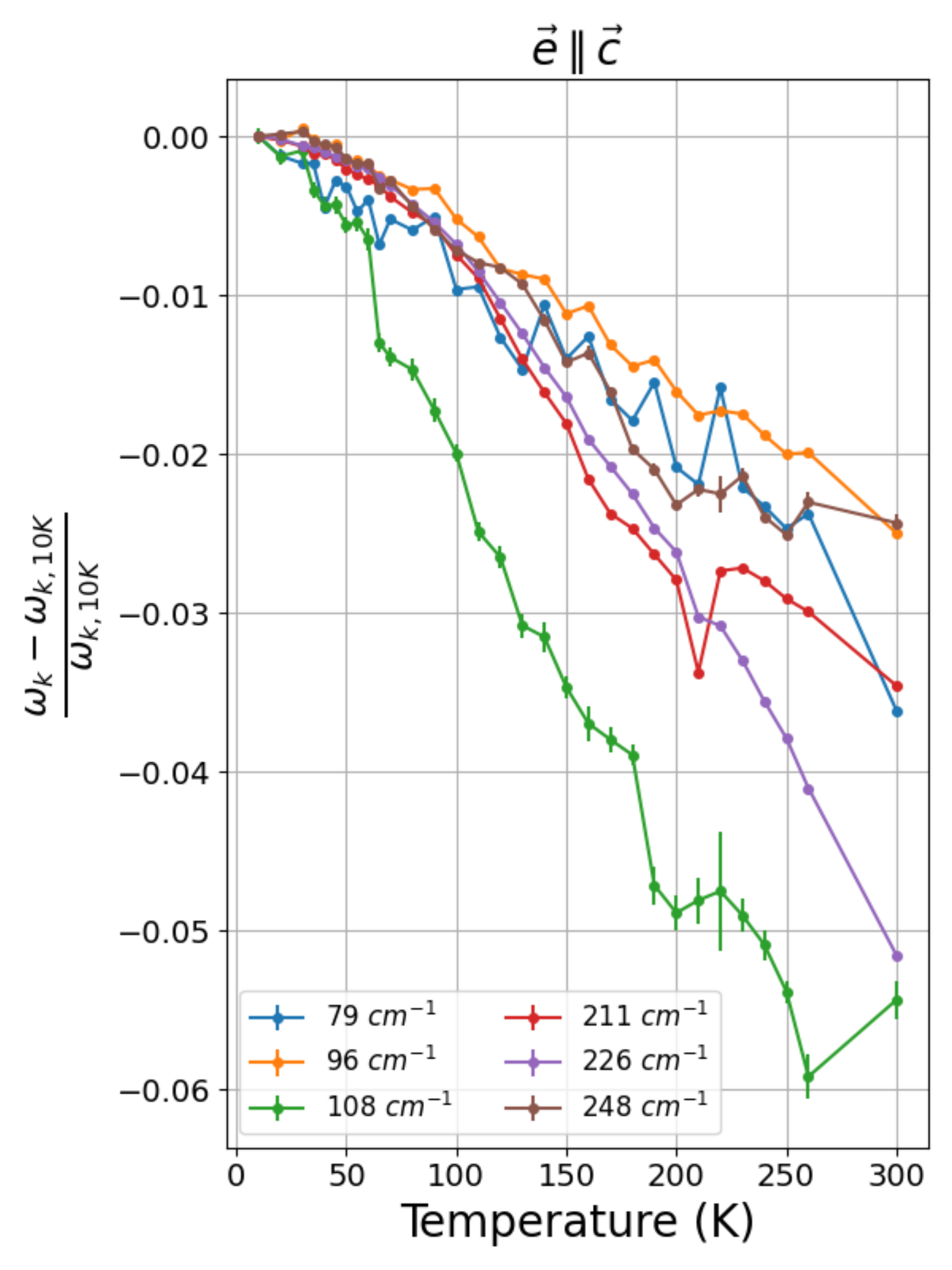}}
  \resizebox{0.32\linewidth}{!}{\includegraphics{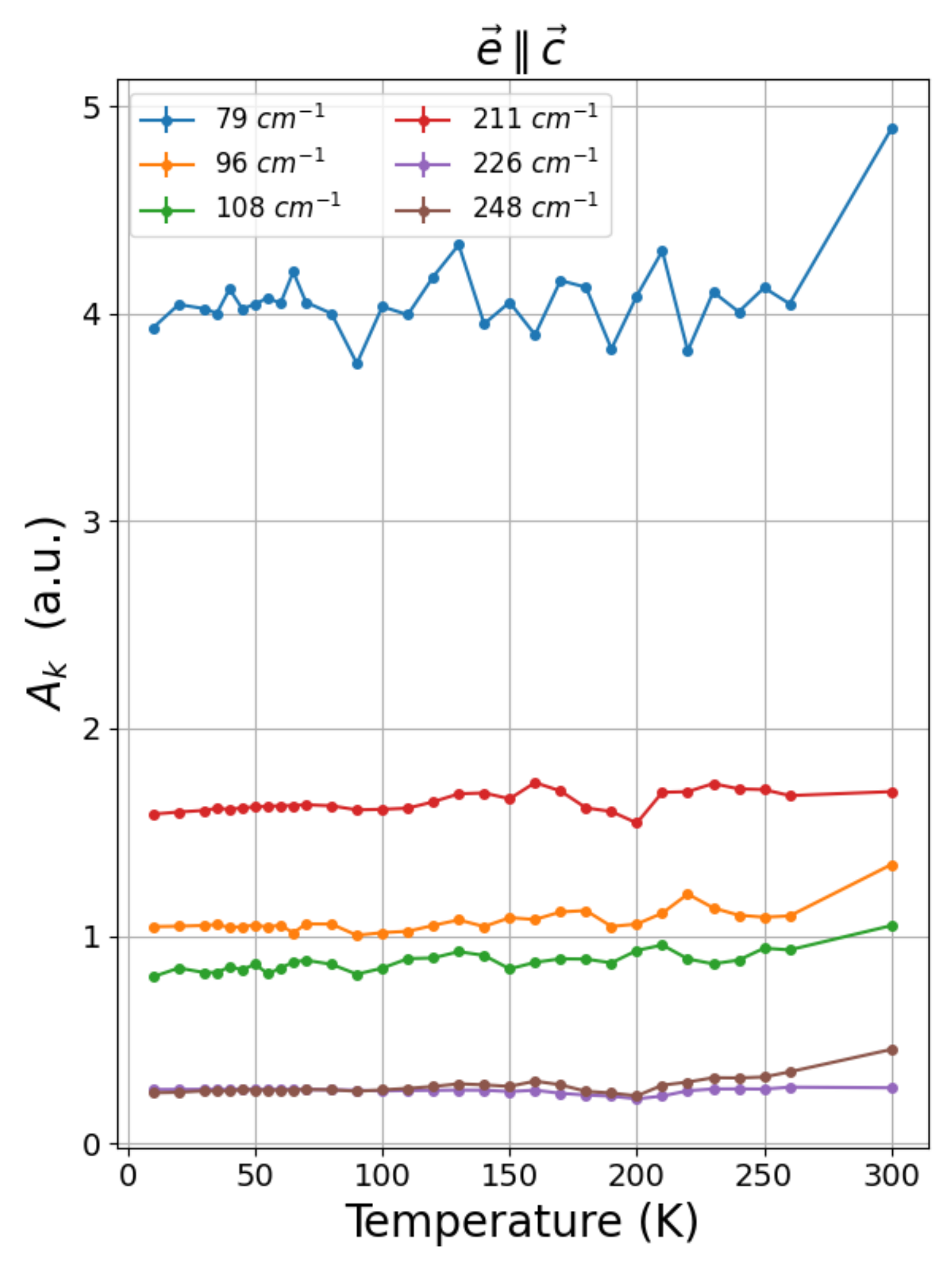}}
  \resizebox{0.32\linewidth}{!}{\includegraphics{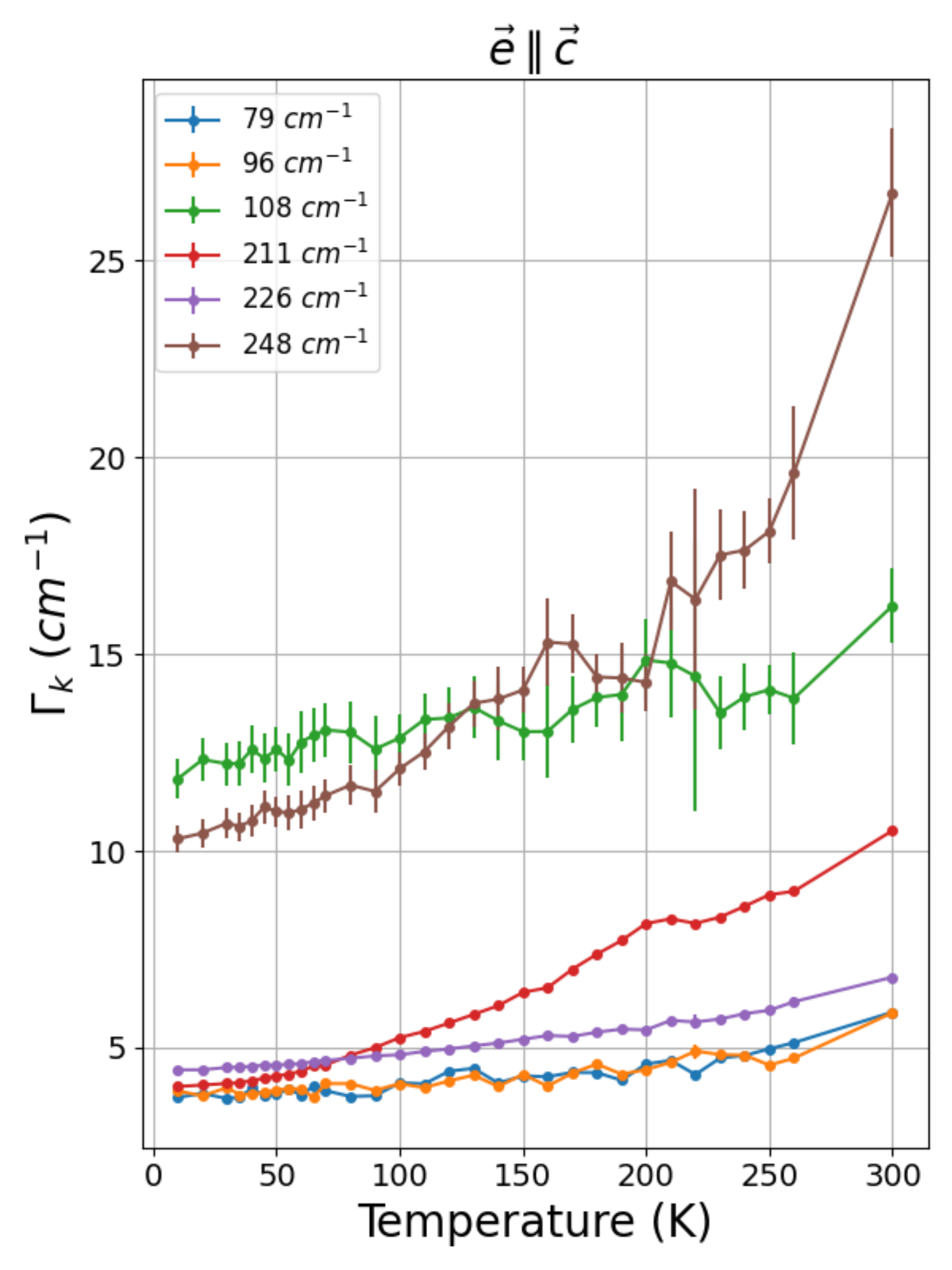}}
  \caption{Temperature dependence of relative variation of $\omega_k$ with respect to low temperature,  $A_k$ and $\Gamma_k$ for $e/\!/ \vec b$ (top panels) and $e/\!/ \vec c$ (bottoms panels).}
  \label{fig:Tdep}
\end{figure*}

The reflectivity measurements of \bfs\ between 10\,K and 300\,K were
performed at quasi-normal incidence, with the electric field along the b-axis
and c-axis in the $Pnma$ standard setting. Figure~\ref{fig:IR10K}a and b
displays the same IR measurements at 10\,K for these two configurations. As we
can see, eight phonons modes are visible when the field is along $\vec b$ and seven
when the field is along $\vec c$. The number of observed modes is unchanged
between 300\,K (see Fig.~\ref{fig:IR300K}) and 10\,K, despite a large
broadening at high temperature, making some phonons difficult to accurately
fit. Experimental phonons frequencies were obtained by fitting the data using
the usual Drude-Lorentz (DL) model for the dielectric function of insulating
materials. The dielectric function is thus expressed as the sum of harmonic
oscillators:
\begin{equation}
\begin{array}{rl}
\displaystyle \epsilon(\omega)= \displaystyle \epsilon_\infty+\sum_{k} \frac{A_k \omega_k^2}{\omega_k^2-\omega^2-i\Gamma_k\omega}
\end{array}
\label{eq_dispersion}
\end{equation}
where $\epsilon_\infty$ is the dielectric constant at infinite frequency,
$\omega_k$ , $A_k$ and $\Gamma_k$ the resonant frequency, the amplitude and
energy-width of the $k^{th}$ harmonic oscillator. The resulting phonon
frequencies are reported for both extreme temperatures (10\,K and 300\,K) in
Table~\ref{tab:expph}, in order to compare them to DFT results. The temperature
dependence of the three parameters, $\omega_k$ , $A_k$ and $\Gamma_k$ are
represented for both polarization configurations in Fig. \ref{fig:Tdep}. Some
of the phonons are too dampened and/or not enough intense to follow accurately
their position, width and amplitude as function of temperature. This is the
case for $161\,\rm cm^{-1}$ ($e/\!/ \vec c$) and $116\,\rm cm^{-1}$
($e/\!/ \vec b$). From the evolution of their energies, we can see that all
phonons undergo a hardening with decreasing temperature. More
interesting, we can observe a quasi-systematic anomaly around 200\,K for each
of these parameters for these phonons.  For example, nearly all phonons
measured for $e/\!/ \vec b$ display a drop in amplitude $A_k$ around 200 K,
concomitant to a change in linewidth evolution $\Gamma_k$. A shift of the
energy $\omega_k$ is also observed at the same temperature for the
$161\,\rm cm^{-1}$ and $241\,\rm cm^{-1}$ phonons. For $e/\!/ \vec c$, mainly
two phonons display a similar anomaly at 200 K in the three parameters
$\omega_k$, $A_k$ and $\Gamma_k$, namely the $211\,\rm cm^{-1}$ and
$248\,\rm cm^{-1}$ ones.  The temperature, where the mentioned phonons behavior
changes, corresponds to the N\'{e}el temperature measured by neutron diffraction
\cite{Zheng} on a sample originating from the very same batch as our
sample. This observation strongly suggest a significant spin-phonon coupling
in this system.


\begin{table}[h]
  \centering
  \begin{tabular}{rr@{\quad}rr}
    \hline \hline  
    \multicolumn{2}{c}{$e/\!/ \vec b$} \rule{0pt}{2.5ex}{} &\multicolumn{2}{c}{$e/\!/ \vec c$}  \\[1ex]
    10\,K & 300\,K & 10\,K & 300\,K \\
    \hline     
    61 & 60   & 79 & 76 \\ 
    75 & 72   & 96 & 94 \\ 
    79 & 78   & 108 & 104 \\ 
    105 & 97 & 163 & - \\ 
    116 & - &  211 & 208 \\ 
    179 & 176   &  226 & 222 \\ 
    210* & 207* & 248 & 246 \\ 
     241 & 236 &  &  \\ 
    \hline \hline 
  \end{tabular}
  \caption{Fitted phonons frequencies ($\rm cm^{-1}$) from infrared
    reflectivity spectra shown in Fig. \ref{fig:IR10K} and \ref{fig:IR300K}
    for both polarization configuration $e/\!/ \vec b$ and $e/\!/ \vec c$. The
    value at 300K of the $116\,\rm cm^{-1}$ ($e/\!/ \vec b$) and
    $161\,\rm cm^{-1}$ ($e/\!/ \vec c$). phonons could not be determined
    within an acceptable error-bar due to loss of intensity and
    dampening. Stars indicate that the value has been graphically estimated.}
  \label{tab:expph}
\end{table}
 
\section{DFT calculations and discussion}
\label{sec:dft}

We first optimized the structure and computed the phonons spectra without spin
polarization, as spins usually induce only small shifts (a few $\rm cm^{-1}$)
of the phonons frequencies. The calculations were performed in the $Pnma$,
$Pmn2_1$ and $Pm$ groups. The optimized geometries can be found in the
supplementary material. 
All three calculations yielded similar energies, within DFT error-bars. Thus, one
can not discriminate between them on this criterium. All optimized
geometries agree well with the experimental single crystal X-ray diffraction
of Ref.~\onlinecite{Krzton_Maziopa}. The comparison (using the amplimode
code~\cite{amplimode2009,amplimode2010}) of the $Pnma$, $Pmn2_1$ and $Pm$
calculations, with the room temperature experimental geometry of
Ref.~\onlinecite{Krzton_Maziopa}, yields similar degrees of lattice distortion
(0.0027) for the three structures and measures of compatibility of 0.028
($Pnma$ and $Pm$) and 0.064 ($Pmn2_1$). 

\subsection{Symmetry Analysis}
69 optical phonon modes and 3 acoustic ones are expected at the $\Gamma$ point. 
Symmetry analysis indicates that the infrared phonons observed with the field
along the $\vec b$ direction should belong to the $B_{3u}$ irreducible representation (irrep) in the $Pnma$
group, and to the $B_{1u}$ irrep when the field is set along the $\vec c$
direction (see character tables in the supplementary material).

The group/subgroup relationships between the different irreps of
the $Pnma$, $Pmn2_1$ and $Pm$ groups are reported on table~\ref{tab:irrep}. They will be
further used for the phonons assignments. 

\begin{table}[h]
  \centering
  \begin{tabular}{c@{\quad}ccll}
    \hline \hline
    $e$&  \multicolumn{2}{c}{$Pnma$} &  \;\;$Pmn2_1$  & \;\;\;\;$Pm$    \\
    \hline \rule{0ex}{3ex}{\hfill}
       & $A_g$ & $\Gamma_1^+$ &  \multirow{2}{*}{$\left. \rule{0ex}{3ex}{\hfill} \right\} A_1\; \Gamma_1$} &
     \multirow{4}{*}{$\left. \rule{0ex}{8ex}{\hfill} \right\} A\; \Gamma_1$}  \\
    $e/\!/ \vec c$ & $B_{1u}$ & $\Gamma_3^-$ & &  \\[1ex]
       & $B_{3g}$ & $\Gamma_2^+$ & \multirow{2}{*}{$\left. \rule{0ex}{3ex}{\hfill} \right\} B_1\; \Gamma_4$} & \\
     $e/\!/ \vec a$  &  $B_{2u}$ & $\Gamma_4^-$ & &  \\[2ex]
       & $B_{1g}$ & $\Gamma_3^+$ & \multirow{2}{*}{$\left. \rule{0ex}{3ex}{\hfill} \right\} A_2\; \Gamma_3$} &
     \multirow{4}{*}{$\left. \rule{0ex}{8ex}{\hfill}\right\} B\; \Gamma_2$} \\
       & $A_u$ & $\Gamma_1^-$ & &   \\[1ex]
       & $B_{2g}$ & $\Gamma_4^+$ & \multirow{2}{*}{$\left. \rule{0ex}{3ex}{\hfill} \right\} B_2\; \Gamma_2$} & \\
     $e/\!/ \vec b$& $B_{3u}$ & $\Gamma_2^-$ & & \\[2.75ex]
    \hline \hline
  \end{tabular}
  \caption{Group/subgroup relationships between the irreducible
    representations of the $Pnma$, $Pmn2_1$ and $Pm$ groups and assignment of the different IR electric
    field polarizations (axes defined as in $Pnma$ standard setting).}
  \label{tab:irrep}
\end{table}

\subsection{The $Pnma$ Space Group}
Out of the 69 optical phonon modes in
the $Pnma$ group, 36  modes are Raman active, 26 modes are IR active.
 The irrep distribution is as follows
\begin{align*}
  & \overbrace{11\, \text{A}_g \oplus 7\, \text{B}_{1g} \oplus 7\, \text{B}_{2g} \oplus 11\, \text{B}_{3g}}^{\text{Raman active}} \\
  \oplus & \overbrace{10\, \text{B}_{1u} \oplus 10\, \text{B}_{2u} \oplus 6\, \text{B}_{3u}}^{\text{IR active}}
            \oplus   \overbrace{7\, \text{A}_u}^{\text{Inactive}}
\end{align*}

The tables of the computed $Pnma$ phonons modes can be found in the supplementary material.

In this group, the number of measured modes exceeds the number of
predicted ones for different irreps. Seven  modes were indeed measured in the
$B_{3u}$ irrep at 300\,K (eight at 10\,K), while only six are predicted by the symmetry analysis.  In
addition, some of the modes cannot be assigned due to the lack of computed
modes in the proper energy range.  In total, there are two $A_g$, three
$B_{3u}$, two $B_{1u}$ and one $B_{2g}$ modes that cannot be assigned within
the $Pnma$ group.  Hence, in agreement with our previous X-ray diffraction
work~\cite{Zheng}, the lattice dynamics  clearly excludes the $Pnma$
group, even in the paramagnetic phase.

\subsection{The $Pmn2_1$ Space Group}
Out of the 69 optical phonon modes expected in the $Pmn2_1$ group,  
all are Raman active and 55 IR active. Their distribution into the group's irreps is the following
$$
\overbrace{\underbrace{21\, \text{A}_1 \oplus 21 \,\text{B}_{1} \oplus 13\,
    \text{B}_{2}}_{\text{IR active}} \oplus 14\, \text{A}_{2} }^{\text{Raman
    active}}
$$

Tables~\ref{tab:Pmn21_res} and~\ref{tab:Pmn21_nonassigned} display the
$Pmn2_1$ computed phonons modes and their best assignment to the measured
modes. The IR modes stem from our measurements at 300\,K, while the
experimental Raman modes are taken from Ref.~\onlinecite{Popovic} at 300\,K.

\setlength\LTleft{0pt}
\setlength\LTright{0pt}
\begin{longtable}[h]{@{\hspace*{0.5ex}}  c d{1} @{\hspace*{2ex}} d{-1}   d{1}d{1} @{\hspace*{5.5ex}}}
  \hline \hline \\[-1.8ex]
  \multicolumn{2}{@{\hspace*{4.5ex}}c}{DFT $Pmn2_1$\hspace*{5ex}}
  &\multicolumn{1}{c}{Raman~\cite{Popovic}\hspace*{5ex}}
  & \multicolumn{2}{@{\hspace*{6ex}}c}{IR~300\,K\hspace*{10ex}}\\
  Irrep  &   \nu (cm^{-1})   &\multicolumn{1}{@{\hspace*{-5ex}}c}{300\,K}& e /\!/{\bf b}  &   e/\!/{\bf c} \\
  \hline \\[-1.8ex] \endfirsthead
  \hline \hline \\[-1.8ex]
  \multicolumn{5}{@{\hspace*{4.5ex}}c}{Continued} \\[1ex] 
  \multicolumn{2}{@{\hspace*{0.5ex}}c}{DFT $Pmn2_1$\hspace*{2ex}}
  &\multicolumn{1}{c}{Raman~\hspace*{5ex}}
  & \multicolumn{2}{@{\hspace*{6ex}}c}{IR~300\,K\hspace*{10ex}}\\
  Irrep  &   \nu (cm^{-1})   &\multicolumn{1}{@{\hspace*{-5ex}}c}{300\,K}& e /\!/{\bf b}  &   e/\!/{\bf c} \\
  \hline \\[-1.8ex] \endhead
  \hline \endfoot
  \endlastfoot
B$_1$  &  29.9   &       &       &         \\ 
A$_1$  &  33.4   &       &       &         \\ 
A$_2$  &  36.5   &       &       &         \\ 
A$_1$  &  39.5   &       &       &         \\ 
B$_1$  &  45.7   &       &       &         \\ 
B$_1$  &  51.5   &       &       &         \\ 
A$_1$  &  52.7   &       &       &         \\ 
B$_1$  &  60.2   &       &       &         \\ 
B$_2$  &  63.7   &       &  60 &         \\ 
A$_1$  &  64.4   & 59.0  &       &         \\ 
A$_2$  &  64.4   &       &       &         \\ 
A$_1$  &  66.6   &       &       &         \\ 
A$_1$  &  67.8   &       &       &     76    \\ 
A$_2$  &  67.9   &       &       &         \\ 
B$_1$  &  70.8   &       &       &         \\ 
B$_2$  &  76.1   &       &  72 &         \\ 
B$_1$  &  79.5   &       &       &         \\ 
A$_2$  &  84.1   &       &       &         \\ 
B$_1$  &  86.4   &       &       &         \\ 
A$_1$  &  87.1   & 88.0  &       &     \\ 
B$_2$  &  89.1   &       &  78 &         \\ 
A$_1$  &  91.9   &       &       &  94  \\ 
B$_1$  &  93.1   &       &       &         \\ 
B$_2$  & 101.1   &       & 97 &         \\ 
A$_2$  & 101.9   &       &       &         \\ 
A$_2$  & 112.2   &       &       &         \\ 
B$_2$  & 112.7   &       & - &         \\ 
A$_1$  & 113.2   & 104.3 &       &  104  \\ 
A$_1$  & 114.7   & 111.0 &       &         \\ 
B$_1$  & 115.4   &       &       &         \\ 
B$_1$  & 117.4   &       &       &         \\ 
A$_1$  & 128.4   & 137.0 &       &         \\ 
B$_2$  & 133.4   &       &       &         \\ 
A$_2$  & 134.3   &       &       &         \\ 
B$_1$  & 135.2   &       &       &         \\ 
A$_1$  & 150.1   &       &       &         \\ 
B$_1$  & 156.2   &       &       &         \\ 
A$_1$  & 160.5   &       &       &         \\ 
B$_1$  & 163.2   &       &       &         \\ 
A$_1$  & 170.4   &       &       &  -  \\ 
B$_1$  & 174.3   &       &       &         \\ 
B$_2$  & 184.1   & 177   & 176 &         \\ 
A$_2$  & 184.5   &       &       &         \\ 
B$_1$  & 186.2   &       &       &         \\ 
B$_2$  & 190.9   &       &       &         \\ 
A$_2$  & 192.3   &       &       &         \\ 
A$_1$  & 195.3   & 195.6 &       &         \\ 
A$_2$  & 224.5   &       &       &         \\ 
B$_2$  & 225.0   & 222.8 &       &         \\ 
A$_1$  & 250.1   &       &       &  246  \\ 
B$_1$  & 251.5   &       &       &         \\ 
A$_2$  & 253.0   &       &       &         \\ 
B$_2$  & 253.2   &       &       &         \\ 
A$_1$  & 256.3   & 267   &       &         \\ 
B$_1$  & 260.9   &       &       &         \\ 
B$_2$  & 260.9   &       &       &         \\ 
A$_2$  & 261.9   &       &       &         \\ 
B$_2$  & 267.7   &       &       &         \\ 
A$_2$  & 267.8   &       &       &         \\ 
A$_2$  & 282.0   &       &       &         \\ 
B$_2$  & 282.5   &       &       &         \\ 
A$_1$  & 295.0   &  280.0&       &         \\ 
A$_1$  & 295.5   &  290.0&       &         \\ 
B$_1$  & 296.0   &       &       &         \\ 
B$_1$  & 298.0   &       &       &         \\ 
B$_1$  & 312.7   &       &       &         \\ 
A$_1$  & 314.8   &       &       &         \\ 
A$_1$  & 321.2   &       &       &         \\ 
B$_1$  & 325.6   &       &       &         \\ 
\hline \hline \\
\caption{Computed phonon modes in the $Pmn2_1$ group, and their best assignment
  to the measured Raman and IR modes (cm$^{-1}$) at 300~K. The DFT
  calculations were carried out without spin polarization.  The Raman modes
  were taken from Ref.~\onlinecite{Popovic} and the IR modes from our
  measurements.}
\label{tab:Pmn21_res} 
\end{longtable}
\FloatBarrier
\begin{table}[h]
  \begin{tabular}{@{\hspace*{5.5ex}}  c @{\hspace*{8ex}} d{-1} @{\hspace*{4ex}}  d{1}d{1} @{\hspace*{6.5ex}}}
  \hline \hline \\[-1.8ex]
  Irrep & \multicolumn{1}{@{\hspace*{0ex}}c}{Raman~\cite{Popovic}\hspace*{5ex}}& \multicolumn{2}{@{\hspace*{6ex}}c}{IR~300\,K\hspace*{13ex}}\\
    & \multicolumn{1}{@{\hspace*{-5ex}}c}{300\,K}     & e /\!/{\bf b}  &   e/\!/{\bf c} \\
  \hline \\[-1.8ex] 
    A$_{1}$ & &   & 208   \\
    A$_{1}$ & &   & 222   \\
    B$_2$  & & 207* & \\
    B$_2$  & & 236  & \\
    \hline \hline
    \end{tabular}
\caption{Measured (300\,K) IR phonon modes (cm$^{-1}$) that could not be assigned to
  computed ones in the $Pmn2_1$ space group. }
\label{tab:Pmn21_nonassigned}
\end{table}
\FloatBarrier

One sees immediately that all Raman modes can be assigned with a good accuracy
in the $Pmn2_1$ group, with an average error of $\simeq 6 \rm \,cm^{-1}$. For
the IR modes two modes are problematic along each direction.

Along the $\vec b$ direction, the modes at $207\,\rm cm^{-1}$ and
$236\,\rm cm^{-1}$ can only be assigned to the computed modes with quite large error
bars (namely $18\,\rm cm^{-1}$ and $17\,\rm cm^{-1}$).  Such error bars are at
the extreme limit of DFT  acceptable error bars, and should attract our
attention. In addition, the mode at $207\,\rm cm^{-1}$ has to be assigned to
the same computed mode as the mode seen in Raman scattering at
222.8~cm$^{-1}$, which is nearly 16~cm$^{-1}$ away. Even if one accept the
large differences between the computed and experimental IR frequencies, the
difference between the Raman and IR measurements seems too large to
be accounted by experimental error bars.

Nevertheless, the main problem occurs for infrared modes when the electric
field is set along the $\vec c$ direction. Indeed, the two most intense modes
(see Fig.~\ref{fig:IR300K}b), at 208\,cm$^{-1}$ and 222\,cm$^{-1}$ are
impossible to assign as there are not any $A_1$ modes in the range
$\rm~195\,cm^{-1}$--$\rm~250\,cm^{-1}$.  
Even if the mode at 208\,cm$^{-1}$
was assigned to the computed mode at $\rm~195\,cm^{-1}$ , with an error of
$\rm~23\,cm^{-1}$, this mode was already assigned to a Raman mode at
$\rm~195.6\,cm^{-1}$.
Once more, the discrepancy between the Raman and IR measured
modes is too large to be accounted by experimental error bars.

Prior to fully exclude the $Pmn2_1$ group, we will have a look at the results
in the $Pm$ group, since recent single crystal X-ray diffraction yielded $Pm$
to be the actual space group in the low temperature magnetic phase \cite{Zheng}.

\subsection{The $Pm$ group}

The 69 optical phonon modes of the Pm space group are distributed into the
following irreps
$$ 42\, \text{A} \oplus 27\, \text{B} $$
All modes are active in Raman and IR.

The table displaying the computed phonons modes, as well as their best
assignment to our IR measurements at 10\,K, and to the measured 20\,K Raman
modes from Ref.~\onlinecite{Popovic}, is provided in supplementary material.
Similar to the $Pmn2_1$ case, all Raman modes can be easily assigned. This is
also the case for the IR modes when the field is along the $\vec b$ direction.
The respective average errors for the Raman and IR $e/\!/\vec b$ are
respectively weaker than 5~cm$^{-1}$ and 6.4~cm$^{-1}$.

Most of the infrared modes with the field along $\vec c$ can also be assigned
with small errors. However, as for the $Pmn2_1$ group, the two most intense
modes (at $211\,\rm cm^{-1}$ and $226\,\rm cm^{-1}$) cannot be properly
assigned. Indeed, the only possibility for the mode at $211\,\rm cm^{-1}$
would be to assign it to the same computed mode as the mode seen at
$200\,\rm cm^{-1}$ in Raman scattering. However a $18\,\rm cm^{-1}$ frequency
difference between Raman and infrared measurements seems quite
unlikely. Regarding the mode at $226\,\rm cm^{-1}$, it seems impossible to
assign it without a significant error ($23\,\rm cm^{-1}$), as there are no
computed modes of the proper symmetry in this energy range.

Looking at the displacement vectors associated with the phonons modes in the
$200$--$300\,\rm cm^{-1}$ range, we see that they are dominated by Fe atoms
displacements. As reported in the literature, neutron pair-distribution
functions clearly show a large magneto-elastic coupling, with Fe atoms
displacements at the inset of the antiferromagnetic (AFM)
order~\cite{Caron2011}. In fact, while the long-range order takes place
between 230\,K and 255\,K according to the authors, short-range magnetic
correlations (with a correlation length of $\xi\sim 35\rm\AA$) are observed
from neutrons diffuse scattering~\cite{Caron2011}, even at room temperature.

As a consequence one may think that, unlike in most systems where the magnetic
structure only shift the phonons modes by a few $\rm cm^{-1}$, in \bfs\ the magnetic
order could have a strong effect on the structure dynamics.

\subsection{Importance of in-ladder magnetic order}
The magnetic propagation vector in the \bfs\ system is
$(1/2,1/2,1/2)$. Computing structure dynamics using the
published~\cite{Liu_2019,Krzton_Maziopa,Popovic,Caron2011} magnetic order
would thus require to use a super-cell of eight unit cells. As stated before,
unfortunately, such a large calculation is beyond our present possibilities.
The magnetic structure is however highly anisotropic, with much larger
magnetic couplings within the ladders than between them. One can thus expect
that 1) it is the intra-ladder magnetic order that is responsible for
the magneto-lattice coupling, and 2) that this effect may also be
present in the paramagnetic phase.  We therefore recomputed the phonons modes,
both in the $Pm$ and $Pmn2_1$ groups, using a double super-cell
($\vec a \times 2 \vec b \times \vec c$), and spin-polarized calculations with
the experimental spin ordering along the ladders direction ($\vec b$).

The first consequence was, as expected, a large energy stabilization
($16.831\,\rm eV$ for the $Pmn2_1$ group and $16.833\,\rm eV$ for the $Pm$
group) compared to the non spin-polarized calculations. The energy difference
between the $Pmn2_1$ and $Pm$ groups remains, however, non significant
($\sim 3\,\rm meV$). Looking at the geometries, one sees that the
spin-polarized optimized geometries differ from the non spin-polarized ones,
as far as the Fe--Fe distances along the ladders are concerned.

\begin{figure}[h]
  \centering
  \resizebox{4cm}{!}{\includegraphics{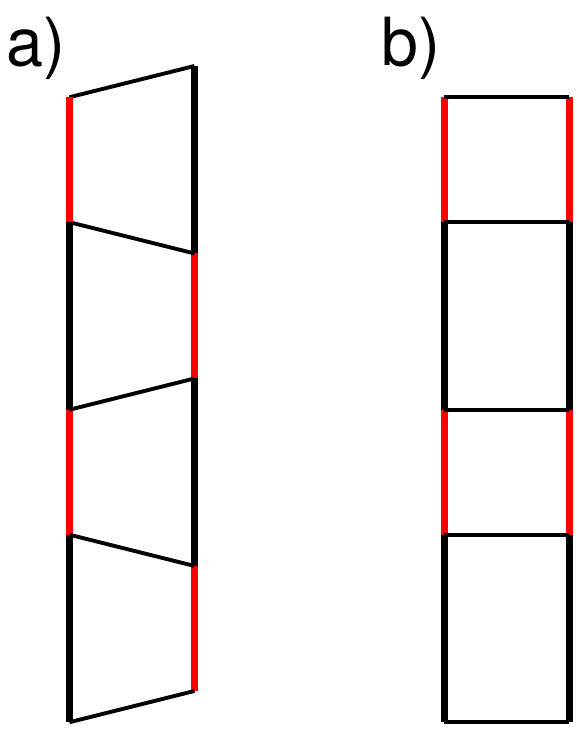}}
  \caption{Schematic representation of the Fe-Fe bond lengths along the
    ladders, a) as obtained in a non-spin polarized geometry optimization, b)
    as obtained when the spin-ordering along the ladders is taken into account.}
  \label{fig:bonds}
\end{figure}
 
The non spin-polarized geometries are all similar to the $Pnma$ experimental
one. Indeed, in all of them we find alternated trapezoidal blocks, as a result
of the Fe--Fe distances alternation along the ladders (see
Fig.~\ref{fig:bonds}~a)). In contrast to this, the spin-polarized optimized
geometries present an alternation between small and large rectangular blocks
(see Fig.~\ref{fig:bonds}~b)), as in the experimental $Pmn2_1$ and $Pm$
structures~\cite{Zheng}.

This feature originates from the spin-ordering along the ladders. This can be
checked by re-optimizing the geometry within a non spin-polarized
calculation, starting from the spin-polarized one. In such calculations the
resulting bond lengths exhibit again trapezoidal blocks
(Fig.~\ref{fig:bonds}~a).

Table~\ref{tab:Pmn21_spins} and~\ref{tab:Pmn21_spins_nonassigned} report the
spin-polarized phonons calculations at the $\Gamma$ point, in the $Pmn2_1$
group. As we worked in a double unit cell, CRYSTAL provides both the $\Gamma$
point and zone-border M-point phonons. We thus wrote a small code to identify
the former, as well as their irrep. For this purpose, we used the usual
group-symmetry projection operator on an irrep $(i)$
$$ P_{(i)} = \frac{1}{|G|} \sum_{g\in G} \chi^*_{(i)}(g)\; g $$
where $|G|$ is the order of the group $G$, $g$ any symmetry operation
belonging to $G$ and $\chi$ the characters.  In the following tables we will
only present the $\Gamma$ point phonons, as they are the only ones seen in
Raman or infrared measurements.

\begin{longtable}{@{\hspace*{4.5ex}}  c d{1} @{\hspace*{6ex}} d{-1} @{\hspace*{6ex}}  d{1}d{1} @{\hspace*{6.5ex}}}
  \hline \hline \\[-1.8ex]
  \multicolumn{2}{@{\hspace*{4.5ex}}c}{DFT $Pmn2_1$\hspace*{5ex}}
  &\multicolumn{1}{@{\hspace*{0ex}}c}{Raman~\cite{Popovic}\hspace*{5ex}}
  & \multicolumn{2}{@{\hspace*{6ex}}c}{IR 300\,K~\hspace*{10ex}}\\
  Irrep  &  \multicolumn{1}{c}{$\nu (\text{cm}^{-1})$}    &\multicolumn{1}{@{\hspace*{-6ex}}c}{300\,K} & e /\!/{\bf b}  &   e/\!/{\bf c} \\
  \hline \\[-1.8ex] \endhead
  \hline \endfoot
  \endlastfoot
 A2 & 24.5 &  &  &  \\ 
 B1 & 25.9 &  &  &  \\ 
 A1 & 28.5 &  &  &  \\ 
 A1 & 34.8 &  &  &  \\ 
 B1 & 38.3 &  &  &  \\ 
 A1 & 38.9 &  &  &  \\ 
 B2 & 47.9 &  &  &  \\ 
 A2 & 49.2 &  &  &  \\ 
 B1 & 50.9 &  &  &  \\ 
 A2 & 52.5 &  &  &  \\ 
 B1 & 54.7 &  &  &  \\ 
 B2 & 59.2 &  & 60 &  \\ 
 A1 & 60.5 & 59 &  &  \\ 
 A2 & 61.2 &  &  &  \\ 
 B2 & 63.1 &  & 72 &  \\ 
 A1 & 66.3 &  &  &  \\ 
 B1 & 68.8 &  &  &  \\ 
 A1 & 69.9 &  &  &  \\ 
 B1 & 72.8 &  &  &  \\ 
 A1 & 80.1 &  &  & 76 \\ 
 B1 & 85.6 &  &  &  \\ 
 B1 & 89.0 &  &  &  \\ 
 A1 & 92.2 & 88 &  &  \\ 
 B2 & 93.1 &  & 78 &  \\ 
 A2 & 94.7 &  &  &  \\ 
 A2 &103.3 &  &  &  \\ 
 B2 &106.0 &  & 97 &  \\ 
 A1 &107.2 & 104.3 &  & 104 \\ 
 A2 &108.5 &  &  &  \\ 
 B2 &109.9 &  & - &  \\ 
 A1 &111.2 & 111 &  &  \\ 
 B1 &111.3 &  &  &  \\ 
 B1 &113.7 &  &  &  \\ 
 A1 &114.6 &  &  &  \\ 
 B1 &125.3 &  &  &  \\ 
 A1 &128.6 &  &  &  \\ 
 B1 &137.6 &  &  &  \\ 
 A1 &144.0 & 137 &  &  \\ 
 B2 &151.3 &  &  &  \\ 
 B1 &152.5 &  &  &  \\ 
 A2 &152.9 &  &  &  \\ 
 B1 &157.1 &  &  &  \\ 
 A1 &162.2 &  &  & - \\ 
 A2 &162.9 &  &  &  \\ 
 B2 &164.1 &  &  &  \\ 
 B2 &185.2 & 177 & 176 &  \\ 
 A2 &185.6 &  &  &  \\ 
 B1 &190.1 &  &  &  \\ 
 A2 &193.8 &  &  &  \\ 
 B2 &194.1 &  &  &  \\ 
 A1 &197.5 & 195.6 &  &  \\ 
 B2 &214.7 &  & 207 &  \\ 
 A2 &214.8 &  &  &  \\ 
 B2 &216.3 & 222.8 &  &  \\ 
 A2 &217.1 &  &  &  \\ 
 A1 &225.1 &  &  & 222 \\ 
 B2 &229.1 &  & 236 &  \\ 
 B1 &231.2 &  &  &  \\ 
 A2 &232.1 &  &  &  \\ 
 A1 &244.6 &  &  & 246 \\ 
 B1 &244.8 &  &  &  \\ 
 B1 &263.9 &  &  &  \\ 
 A1 &265.4 & 267 &  &  \\ 
 B1 &283.3 &  &  &  \\ 
 A1 &285.1 & 280 &  &  \\ 
 A1 &288.8 &  &  &  \\ 
 B1 &290.7 &  &  &  \\ 
 A1 &295.4 & 290 &  &  \\ 
 B1 &299.4 &  &  &  \\ 
 \hline \hline \\
 \caption{Measured and computed phonon modes. The DFT calculations were
   carried out in the Pmn2$_1$ space group. The Raman modes were taken
   from~\cite{Popovic} (measurements at 300\,K). The IR modes were taken from
   our measurements at 300\,K.}
\label{tab:Pmn21_spins}	
\end{longtable}
\begin{table}[h]
  \begin{tabular}{@{\hspace*{4.5ex}}  c @{\hspace*{6ex}} d{-1} @{\hspace*{6ex}}  d{1}d{1} @{\hspace*{6.5ex}}}
  \hline \hline \\[-1.8ex]
  Irrep & \multicolumn{1}{@{\hspace*{0ex}}c}{Raman~\cite{Popovic}\hspace*{5ex}}& \multicolumn{2}{@{\hspace*{6ex}}c}{IR 300\,K~\hspace*{10ex}}\\
&\multicolumn{1}{@{\hspace*{-6ex}}c}{300\,K} & e /\!/{\bf b}  &   e/\!/{\bf c} \\
    \hline \\[-1.8ex]
    B$_{2}$ & & 78  &    \\
    A$_{1}$ & &   & 94   \\
    A$_{1}$ & &   & 208   \\

    \hline \hline
    \end{tabular}
    \caption{Measured (300\,K) IR phonon modes (cm$^{-1}$) that could not be
      assigned to computed ones in the $Pmn2_1$ space group. }
\label{tab:Pmn21_spins_nonassigned}
\end{table}
\FloatBarrier

As can be seen, there still are three infrared phonons modes that cannot be
assigned with the computed frequencies. The $B_2$ mode at 78\,cm$^{-1}$ and
the A$_{1}$ modes at 94\,cm$^{-1}$ and 208\,cm$^{-1}$. In fact, the two A$_{1}$
modes could be assigned to computed ones with reasonable errors. However, as
previously, there are Raman modes already associated with these frequencies
(namely at 88\,cm$^{-1}$ and 196\,cm$^{-1}$) and the differences between the
experimental modes are too large to be accounted for experimental error bars.
The $Pmn2_1$ space group is thus
unlikely to be \bfs\ space group, even at 300~K.

Let us now check the $Pm$
group. Table~\ref{tab:Pm_spins} reports the spin-polarized phonons
calculations at the $\Gamma$ point, in the $Pm$ group.
\begin{longtable}{@{\hspace*{0.0ex}}  c d{1} @{\hspace*{3ex}} d{-1}d{-1} @{\hspace*{2ex}}
    d{1}d{1}@{\hspace*{2ex}} d{1}d{1} @{\hspace*{0.5ex}}}
  \hline \hline \\[-1.8ex]
  \multicolumn{2}{@{\hspace*{0.ex}}c}{DFT\hspace*{1ex}}
  &\multicolumn{2}{@{\hspace*{0ex}}c}{Raman~\cite{Popovic}\hspace*{1ex}}
  & \multicolumn{2}{@{\hspace*{0ex}}c}{IR 300\,K~\hspace*{0ex}}
  & \multicolumn{2}{@{\hspace*{0ex}}c}{IR 10\,K~\hspace*{2ex}}\\
  Irrep  &  \multicolumn{1}{c}{$\nu (\text{cm}^{-1})$}
  &\multicolumn{1}{c}{300\,K}  &\multicolumn{1}{c}{20\,K\hspace*{2ex}}
  & e /\!/{\bf b}  &   e/\!/{\bf c} & e /\!/{\bf b}  &   e/\!/{\bf c} \\
  \hline \\[-1.8ex] \endhead
  \hline \endfoot
  \endlastfoot
   B  &      24.2  &    &    &    &    &    &    \\  
   A  &      26.2  &    &    &    &    &    &    \\  
   A  &      28.7  &    &    &    &    &    &    \\  
   A  &      34.9  &    &    &    &    &    &    \\  
   A  &      38.5  &    &    &    &    &    &    \\  
   A  &      39.2  &    &    &    &    &    &    \\  
   B  &      49.0  &    &    &    &    &    &    \\  
   B  &      49.7  &    &    &    &    &    &    \\  
   A  &      51.1  &    &    &    &    &    &    \\  
   B  &      53.5  &    &    &    &    &    &    \\  
   A  &      54.6  &    &    &    &    &    &    \\  
   B  &      59.7  &    &    &  60  &    &  61  &    \\  
   A  &      60.6  &  59  &  63.4  &    &    &    &    \\  
   B  &      64.0  &    &    &  72  &    &  75  &    \\  
   A  &      66.3  &    &    &    &    &    &    \\  
   B  &      67.7  &    &    &  78  &    &  79  &    \\  
   A  &      69.0  &    &    &    &    &    &    \\  
   A  &      70.1  &    &    &    &    &    &    \\  
   A  &      72.9  &    &    &    &    &    &    \\  
   A  &      79.9  &    &    &    &  76  &    & 79  \\  
   A  &      85.8  &    &    &    &    &    &    \\  
   A  &      89.3  &  88  &  89  &    &    &    &    \\  
   A  &      92.3  &    &    &    &  94  &    &  96  \\  
   B  &      93.1  &    &    &    &    &    &    \\  
   B  &      94.8  &    &    &    &    &    &    \\  
   B  &     105.0  &    &    &  97  &    &  105  &    \\  
   B  &     106.2  &    &    &    &    &    &    \\  
   B  &     106.7  &    &    &    &    &    &    \\  
   A  &     107.3  &  104.3  &  108  & 104  &    &  108  \\  
   B  &     108.8  &    &    & -  &    &  116  &    \\  
   A  &     111.4  &    &    &    &    &    &    \\  
   A  &     111.4  &    &    &    &     &    &    \\  
   A  &     113.7  &    &    &    &    &    &    \\  
   A  &     114.6  &  111  &  115  &    &    &    &    \\  
   A  &     125.2  &    &    &    &    &    &    \\  
   A  &     128.4  &    &    &    &    &    &    \\  
   A  &     137.6  &    &    &    &    &    &    \\  
   A  &     143.9  &  137  &  143  &    &    &    &    \\  
   B  &     151.4  &    &    &    &    &    &    \\  
   A  &     152.5  &    &    &    &    &    &    \\  
   B  &     152.8  &    &    &    &    &    &    \\  
   A  &     157.0  &    &    &    &    &    &    \\  
   A  &     162.1  &    &    &    & - &    &  163  \\  
   B  &     164.1  &    &    &    &    &    &    \\  
   B  &     165.3  &    &    &    &    &    &    \\  
   B  &     176.3  &  177  &  183.8  &  176  &    &  179  &    \\  
   B  &     185.7  &    &    &    &    &    &    \\  
   B  &     186.5  &    &    &    &    &    &    \\  
   B  &     188.8  &    &  198  &    &    &    &    \\  
   A  &     190.3  &    &    &    &    &    &    \\  
   A  &     197.5  &  195.6  &  200  &    &    &    &    \\  
   B  &     210.2  &    &    &  207*  &    &  210*  &    \\  
   B  &     211.9  &    &    &    &    &    &    \\  
   B  &     216.0  &    &    &    &    &    &    \\  
   B  &     217.1  &    &    &   &    &   &    \\  
   A  &     225.6  &    &    &    &  208  &    &  211  \\  
   B  &     230.7  &  222.8  &  228  &    &    &    &    \\  
   A  &     231.2  &    &    &    &  222  &    &  226  \\  
   B  &     232.9  &    &    &  236  &    &  241  &    \\  
   A  &     245.0  &    &    &    &    &    &    \\  
   A  &     245.3  &    &    &    &  246  &    &  248  \\  
   A  &     264.3  &    &    &    &    &    &    \\  
   A  &     265.6  &  267  &  272  &    &    &    &    \\  
   A  &     283.7  &  280  &  288.7  &    &    &    &    \\  
   A  &     285.6  &    &    &    &    &    &    \\  
   A  &     288.9  &    &    &    &    &    &    \\  
   A  &     291.0  &    &    &    &    &    &    \\  
   A  &     295.6  &  290  &  296.5  &    &    &    &    \\  
   A  &     299.5  &    &    &    &    &    &    \\  
  \hline \hline \\
  \caption{Spin-polarized calculation of \bfs\ phonon modes in the $Pm$ group
    and their best assignment to the experimental Raman~\cite{Popovic} and
    infrared modes. Modes with stars are estimated graphically, thus they should
    be taken with caution.}
\label{tab:Pm_spins}	
\end{longtable}
\FloatBarrier

As can be seen in table~\ref{tab:Pm_spins}, all experimental Raman and
infrared modes can now be easily assigned to the theoretical ones. The average
error on the Raman modes at 20 K and 300 K is smaller than 4\,cm$^{-1}$, while it is smaller
than 6\,cm$^{-1}$ for the infrared modes measured at 300\,K and 5\,cm$^{-1}$
for the infrared modes measured at 10\,K. From these results one can thus
confirm that the space group of \bfs\ is $Pm$ in the magnetic phase, and infer it
should also be $Pm$ and not $Pmn2_1$ or $Pnma$ at 300\,K, in the paramagnetic
phase.

\section{Conclusion} 
\label{sec:Conclusion}
In the present paper we propose an experimental and theoretical study of the
structure dynamics of \bfs. The comparison of our infrared measurements and
computed phonons frequencies in the $Pnma$, $Pmn2_1$ and $Pm$ space groups
shows that $Pm$ is the only possible space group, not only in the low
temperature magnetic phase, but also on the high temperature paramagnetic
phase.

Our calculations reveals that the magnetic order within the ladders plays
a crucial role, even in the paramagnetic phase. This is in agreement with neutrons
diffuse scattering experiments, that support the existence of a short range
magnetic order in the paramagnetic phase, with an approximate correlation
length of 35\AA~\cite{Caron2011}. In fact, both the optimized structure and the
phonons modes strongly differ in some key-points, when the magnetic order along
the ladders is taken into account in the calculations. The first difference
resides in the Fe-Fe distances along the ladder legs. Indeed, in a non
spin-polarized calculation, the Fe-Fe bond lengths along the ladders exhibit a
short/long long/short trapezoidal pattern (see figure~\ref{fig:bonds}a));
whereas in a spin-polarized calculation, the spin-ordering along the ladders
induces a complete change of the Fe-Fe bond lengths pattern, with an
alternation of large and small rectangular blocks (see
figure~\ref{fig:bonds}b)). The latter blocks pattern is in good agreement with
our X-ray diffraction measurements of Ref.~\onlinecite{Zheng}, while the
trapezoidal pattern is in good agreement with the $Pnma$ X-ray structure from
Ref~\onlinecite{Caron2011}. It is also of crucial importance to note that the
rectangular blocks pattern is fully compatible with the block
antiferromagnetic structure, while the trapezoidal pattern is not.  Indeed,
if the short/long Fe-Fe bonds are associated to AFM/FM exchange
integrals, the experimental magnetic ordering seen along the ladders becomes
obvious. The determination of the magnetic integrals associated with the Pm
structure should thus be our next piece of work.  This resolves the puzzle of
the crystal structure incompatibility with the magnetic structure\vio{,} often
discussed in the literature.

At 660\,K \bfs\ undergoes a $Cmcm$ to $Pnma$ phase
transition~\cite{Svitlyk2013}, followed by another transition at 425\,K. This
transition was shown, by scanning transmission electron microscopy, to be
associated with an in-ladder tetramerisation, a differentiation between the
two ladders of the unit cell and a room temperature
polarization~\cite{Du2020}. Putting these experimental results into
perspective with our structure dynamic study, one can built a coherent picture
for the \bfs\ phase diagram.  After the 660\,K phase transition from the
$Cmcm$ group to $Pnma$, the 425~K phase transition can be associated to a
$Pnma$ to $Pm$ symmetry lowering\vio{,} induced by the inset of the short-range
magnetic correlations within the ladders. Such a transition would be in
agreement with the inset of the tetramerisation, the polarization and the release
of the symmetry relationship between the two ladders. It would also account
for the observed strong coupling between the lattice and the short-range magnetic order.

\section{Acknowledgment}
The theoreticians among the
authors thank the IDRIS (project n$^\circ$91842) and CRIANN (project
n$^\circ$2007013) computer centers on which the calculations presented in this
paper have been done.
The experimental  work was financially supported by the ANR COCOM 20-CE30-0029 and
by the CSC scholarship (No. 201806830111). We also thank SOLEIL for
synchrotron beam time (Proposal 20211300).

\bibliographystyle{apsrev4-2} \bibliography{Refs}

\end{document}